\documentstyle[12pt,epsfig]{article} 
\newcommand{\be}{\begin{equation}}   
\newcommand{\ee}{\end{equation}}   
\newcommand{\bear}{\begin{eqnarray}}   
\newcommand{\eear}{\end{eqnarray}}   
\newcommand{\ba}{\begin{array}}   
\newcommand{\ea}{\end{array}}

\newskip\humongous \humongous=0pt plus 1000pt minus 1000pt

\newif\ifdtup

\def\oldreffmt#1{\rlap{[#1]} \hbox to 2\parindent{}}

\def\figfmt#1{\rlap{Figure {#1}} \hbox to 1in{}}   
   
\def\Tr{\mathop{\rm Tr}}

\def\VEV#1{\left\langle #1\right\rangle}

\def\beq{\begin{equation}}   
\def\eeq{\end{equation}}   
\def\bea{\begin{eqnarray}}   
\def\eea{\end{eqnarray}}   
\def\half{\frac{1}{2}}   
   
\def\bq{\begin{quote}}   
\def\eq{\end{quote}}

\def\half{\frac{1}{2}}        
\def \lta {\mathrel{\vcenter   
     {\hbox{$<$}\nointerlineskip\hbox{$\sim$}}}}   
    
\relax   

\newdimen\tdim   
\tdim=\unitlength   
\def\bar{\overline}

\begin{document}

\pagestyle{empty}
\begin{titlepage}
\def\thepage {}    

\title{  \bf  
Topological Solitons\\
from DeConstructed Extra Dimensions } 
\author{
\bf  Christopher T. Hill\\
{\small {\it Fermi National Accelerator Laboratory}}\\
{\small {\it P.O. Box 500, Batavia, Illinois 60510, USA}}
\thanks{e-mail: 
hill@fnal.gov  }
}

\date{August 20, 2001}

\maketitle

\vspace*{-10.5cm}
\noindent

\begin{flushright}
FERMILAB-Pub-01/252-T \\ [1mm]
Sept. 7, 2001
\end{flushright}

\vspace*{14.1cm}
\baselineskip=18pt

\begin{abstract}

  {\normalsize
A topological monopole-like
field configuration exists for Yang-Mills
gauge fields in a $4+1$ dimensions.  When  
the extra dimension is compactified to $3+1$ dimensions
with periodic lattice boundary conditions, 
these objects
reappear in the low energy effective theory
as a novel solution, a gauged-bosonic Skyrmion. When
the low energy theory spontaneously breaks,
the Nambu-Goldstone mode develops a VEV, and the gauged-bosonic 
Skyrmion morphs into a `t Hooft--Polyakov monopole.
} 
\end{abstract}

\vfill
\end{titlepage}

\baselineskip=18pt
\renewcommand{\arraystretch}{1.5}
\pagestyle{plain}
\setcounter{page}{1}


\section{Introduction}

This is a tale of three well-known topological
solitons: the instanton, the Skyrmion, and the `t Hooft-Polyakov
monopole. All three of these objects arise from a common source
when Yang-Mills fields propagate in a $4+1$ bulk compactified by
periodic boundary conditions to $3+1$ dimensions. A
consistent description of this dimensional descent is 
most readily obtained through deconstruction, or
latticization, of the extra compactified dimension.
The structures of the conserved Chern-Simons currents 
neatly match, as they must, between the effective descriptions.

We start in $4+1$ dimensions with an $SU(2)$ Yang-Mills
theory and note that  there
are ``instantonic monopoles'' (IM). These are static, topologically
stable solutions of the pure Yang-Mills gauge theory and
represent nontrivial homotopy of $\Pi_3(SU(2))$, the
winding of the field configuration on the surface $S_3$
at infinity in four spatial dimensions.
These objects were considered about a year ago by
Ramond and the present author \cite{ramond, ramond2}, and
they are evidently the anticipated pure-Yang-Mills solitons that
can exist only in $4+1$ by Deser \cite{deser}. These are
essentially instantons \cite{belavin, thooft}
``lifted'' to become the spatial
configurations of a static object. 
For the Instantonic Monopole  we can choose in $4+1$
the (noncompactified) vector potentials
(where $A,B,..$ run from $0$ to $4$, $x^4$ is our 5th dimension; time
is $x^0$):
\be
A_0^a = 0 
\qquad\qquad
A^a_4 \frac{\tau^a}{2} = -\frac{1}{\bar{g}}\frac{\vec{x}\cdot \tau }{\lambda^2 + r^2} 
\qquad\qquad
A^a_i \frac{\tau^a}{2} = 
\frac{1}{\bar{g}}\frac{(x_4 \tau_i + \vec{x}_j\epsilon^{ijk}\tau_k) }{\lambda^2 + r^2} 
\ee
This field configuration
has an associated conserved
topological current \cite{ramond}:
\be
\label{cur1}
Q_A = \frac{\bar{g}^2}{16\pi^2}\epsilon_{ABCDE}\Tr(F^{BC}F^{DE})
\ee
The resulting field strength is self--dual
as a static configuration, i.e., $F_{AB}=\tilde{F}_{0AB}$. 
It has a mass given by $8\pi^2/\bar{g}^2$
where $\bar{g}$ is a $4+1$ coupling constant with dimension
(mass)$^{-1/2}$. This mass is essentially $M_{KK}/\alpha$ where $M_{KK}$
is the lowest KK-mode mass when the theory is compactified.

If we compactify the 5th dimension and ``deconstruct,'' or latticize
the compactified dimension, 
we obtain an equivalent low
energy effective theory in
$3+1$ dimensions \cite{wang0,wang1,georgi}. 
With periodic boundary conditions in our compactification,
the $A^a_4$ vector potential becomes a Nambu-Goldstone 
zero mode, and the product
of Wilson links in the $x^4$ dimension becomes a 
low energy chiral field $U$, the exponentiated
Nambu-Goldstone zero mode.  Keeping only
a single lattice brane as an approximation, the effective
low energy $3+1$ theory is then the gauged chiral
Lagrangian:
\be
\label{eq1}
L = \half v^2\Tr [D_\mu, U^\dagger][D^\mu, U] -\half \Tr F_{\mu\nu}F^{\mu\nu}
\ee
where $F_{\mu\nu}\equiv F_{\mu\nu}^a\tau^a/2$, and
with $U=\exp(i\phi/v)$, $\phi=\phi^a\tau^a/2$, where
$\phi$ is essentially the Wilson line $\sim ig\int dx^4 A^4$
in $4+1$ with $A_\mu = A_\mu^a\tau^a/2$. 
 The action of the covariant derivative upon $U$ is
$[D_\mu, U] =\partial_\mu U -ig[A_\mu, U]$. 

What then is the fate of the instantonic monopole
in this low energy theory viewed as a
dimensional deconstruction?  In
an attempt to clarify what the instantonic monopole is
in a compactified theory, the present author 
was originally
motived to consider latticizing 
the extra dimension \cite{wang0,wang1}.
As we will see, a remarkable correspondence emerges.

The first problem is to ask whether a compactified
IM solution exists?  This question was approached in ref.\cite{ramond},
but see \cite{ramond2}.  One employs the
method of images and exact multi-instanton solutions
to construct a solution satisfying the periodic boundary conditions.
This is well known from  
finite temperature applications of instantons
 \cite{yaffe}.  
For compactification
with periodic boundary conditions,
the low energy pseudoscalar $A_4^a$
remains as a zero mode, while
with orbifold boundary conditions this mode is absent.
Correspondingly, while it is straightforward 
to compactify the IM with periodic 
boundary conditions, it is not with the orbifold boundary
conditions. This is 
a consequence of topology; the topology is determined by the winding of
field $U = \exp(ig\int dx^4 A_4)$ throughout the manifold on large
distances, requiring the $A_4$ zero mode.
This will form the basis of the correspondence with a low
energy effective Lagrangian description below.

One important consequence of compactification of the IM
is the following \cite{yaffe}. 
In an infinite bulk the IM is conformally
invariant. The solution has a scale parameter $\lambda$ but
the action is independent of $\lambda$. The action
density is concentrated in an arbitrarily
large region $r \lta \lambda$, ergo arbitrarily large 
instantons exist.
When a dimension is compactified with a length scale $\delta$,
however,
the field strength configuration changes, and the action
density has appreciable values only over
$r \lta \delta $. Hence, compactification effectively
cuts-off the large instantonic monopoles and gives
them a size of order the compactification scale. 

For the effective description of the $4+1$ IM
in the $3+1$ effective Lagrangian we note that
the theory of eq.(\ref{eq1}), which
is just a conventional gauged chiral Lagrangian,
does indeed contain a
novel soliton, a ``bosonic gauged Skyrmion.'' This, we will
argue,
is the $3+1$ correspondence of the instantonic monopole
of $4+1$.
This object is an ``inverted 
Skyrmion'' built out of the 
$\exp(i\phi/v)$ Wilson link chiral field. At infinity
$\phi/v\rightarrow \pi \hat{x}\cdot\vec{\tau}$
is a hedgehog, while at the origin
$\phi/v\rightarrow 0$. This is inverted
from the usual Skyrmion, but still trivially
represents the nontrivial $\Pi_3(SU(2))$ mapping into the
$3+1$ spatial volume (which, of course, corresponds to the
spatial $S_3$ surface of $4+1$). 
There are, however, other key differences between the Bosonic
Gauged Skyrmion {BGS} and the usual Skyrmion.

The usual Skyrmion
has a nontrivial Wess-Zumino (WZ) term which gives it 
unusual spin and statistics. Choosing the quantized
WZ term coefficient to match to $N_c=3$ QCD, the WZ term
makes the Skymion into a spin$-\half$ baryon. In
the present case the gauging by $SU(2)$ forbids the WZ
term, and the gauged Skyrmion is a bosonic object of spin$-0$.

The usual Skyrmion carries a nontrivial topological
charge determined from a Chern-Simons current.
This current is nontrivially modified in the
present case, and is seen to involve a new term
which  matches the current of eq.(\ref{cur1}) 
under dimensional descent.

\section{Gauge Invariant Chern-Simons Current}
 
The usual Skyrmion is associated with the
conserved, normalized Chern-Simons current, and carries
a unit charge:
\be
Q^\mu = \frac{1}{24\pi^2}\epsilon^{\mu\nu\rho\sigma}\Tr \left(  
U^\dagger(\partial_\nu U)
U^\dagger(\partial_\rho U) U^\dagger(\partial_\sigma U) \right)
\ee
The index for the usual Skyrmion anzatz, $U= \cos(f(r)) + 
\vec{\hat{x}}\cdot\tau \sin(f(r))$, is then
\be
\int d^3x \frac{1}{24\pi^2}\epsilon^{ijk}\Tr \left(  
U^\dagger\partial_i U 
U^\dagger \partial_j U U^\dagger\partial_k U \right)
= \frac{1}{2\pi}[2(f(\infty) -f(0)) + \sin( 2f(\infty))
-\sin( 2f(0))]
\ee
$f(r) = \phi(r)/v$ is a kink-like
configuration that 
runs from $f(0)=0$ to $f(\infty)=\pi$, and
thus has unit charge. Note that $f\rightarrow \pm f + N\pi$
is a discrete symmetry (with charge conjugation), 
so the usual QCD Skyrmion with
$f(0)=\pi$ to $f(\infty)=0$ is equivalent. 

When we go over to the gauged case, we might
guess that the gauge
invariant generalization of the Chern-Simons current is:
\be
Q_1^\mu = \frac{1}{24\pi^2}\epsilon^{\mu\nu\rho\sigma}\Tr \left(  
U^\dagger [D_\nu, U] 
U^\dagger [D_\rho, U] U^\dagger [D_\sigma, U] \right)
\ee
However, $Q_1$ is not conserved, as seen by explicit calculation:
\bea
\label{eqnc}
\partial_\mu Q_1^\mu & = &
\frac{ig}{16\pi^2}\epsilon^{\mu\nu\rho\sigma}\Tr \left( F_{\mu\nu}
[D_\rho, U][ D_\sigma, U^\dagger] -  
F_{\mu\nu}[D_\rho, U^\dagger][ D_\sigma, U] \right)
\eea
Note as a technical
aside that $Q_1^\mu=V+A$ is constructed from purely
right-handed (or left-handed) chiral currents. One cannot build 
a conserved Chern-Simons current
out of the product of mixed
vector $V=(1/2)(U^\dagger [D_\rho, U] + U [D_\rho, U^\dagger])$
and axial vector $A=(1/2)(U^\dagger [D_\rho, U] - U [D_\rho, U^\dagger])$
currents, even in the ungauged case.
$Q_1^\mu$ transforms, however, as a vector
under parity, since  $\phi \rightarrow -\phi$
hence $U\leftrightarrow U^\dagger$, $\epsilon^{\mu\nu\rho\sigma} 
\leftrightarrow -\epsilon^{\mu\nu\rho\sigma}$ and $D_\mu \rightarrow - D_\mu$,
so $Q^\mu\rightarrow -Q^\mu$.  Note also,
\be
Q_1^\mu = -\frac{1}{24\pi^2}\epsilon^{\mu\nu\rho\sigma}\Tr \left(  
U [D_\nu, U^\dagger] 
U [D_\rho, U^\dagger] U [D_\sigma, U^\dagger] \right)
\ee
using $U^\dagger [D_\nu, U] = -[D_\nu, U^\dagger ] U$
and cyclicity of the trace, hence $Q_1$ is equivalent to
a current built out of pure left-handed chiral currents,
i.e., the Chern-Simons current is unique.

Does there exist a conserved current to match to the
$4+1$ conserved current of  eq.(\ref{cur1})? 
With the nontrivial gauge fields we can 
presently introduce two new currents:
\be
\label{cur2}
Q_2^\mu  = \frac{ig}{16\pi^2}\epsilon^{\mu\nu\rho\sigma}\Tr \left(  
F_{\nu\rho} U^\dagger [D_\sigma, U] 
- F_{\nu\rho} U[D_\sigma, U^\dagger ]\right)
\ee
\be
\label{cur3}
Q_3^\mu  =  \frac{ig}{16\pi^2}\epsilon^{\mu\nu\rho\sigma}\Tr \left(  
F_{\nu\rho} U^\dagger [D_\sigma, U] 
+ F_{\nu\rho} U[D_\sigma, U^\dagger ]\right)
\ee
These latter currents are indeed expected to play a role
in the matching because in $4+1$
dimensions we had the conserved current of
the instantonic-monopole:
\be
\epsilon_{ABCDE}\Tr(F^{BC}F^{DE}) \sim \epsilon^{\mu\nu\rho\sigma}\Tr( F_{\mu\nu}
U^\dagger [D_\sigma, U])
\ee
and $U^\dagger [D_\sigma, U]\sim F_{\sigma 4}$ is the 
appropriate dimensional descent correspondence of $A_4$ to
the Nambu-Goldstone boson.

We see that $Q_2$ has normal vectorial parity, 
and it can thus form a vector combination with
$Q_1$.
$Q_3$ is an axial vector under parity.
Computing the divergence of $Q_2$ we obtain
the opposite of the {\em rhs} eq.(\ref{eqnc}), and
we thus arrive at the conclusion that
there is a new conserved current:
\be
\label{eqb}
\widetilde{Q}^\mu = Q_1^\mu + Q_2^\mu  \qquad\qquad
\partial_\mu\left( \widetilde{Q}^\mu \right) = 0
\ee
$\widetilde{Q}^\mu$ is thus the $3+1$ current corresponding
to the $4+1$ eq.(\ref{cur1}) under dimensional descent.
The new index remains an
exact differential and is discussed below in eq.(\ref{eqxx}).

For completeness, notice that
the axial current is not conserved:
\bea
\partial_\mu Q_3^\mu 
& = &
-\frac{ig}{16\pi^2}\epsilon^{\mu\nu\rho\sigma}\Tr \left(  
 F_{\mu\nu}[D_\rho, U] [D_\sigma, U^\dagger] 
+ F_{\mu\nu} [D_\rho, U^\dagger][D_\sigma, U] \right)
\nonumber \\
& & +\; \frac{ig}{16\pi^2}\epsilon^{\mu\nu\rho\sigma}\Tr \left(  
 F_{\mu\nu}U^\dagger F_{\rho\sigma}U -  F_{\mu\nu}F_{\rho\sigma}\right)
\eea
The latter term resembles an anomaly.  
If, for the usual Skyrmion, we gauged only the left-handed
or right-handed pieces of $U$, e.g.,
as in the electroweak theory, then we would obtain
the normal current algebra anomalies through
these manipulations, but our baryon number
current must be modified as in eq.(\ref{eqb}). 

\section{Energetics}

The existence of the 
conserved current $\widetilde{Q}^\mu$ guarantees that
there are nontrivial Skyrmionic configurations including
the gauge fields. The core profile of the Skyrmion must,
moreover,  act
as a source to Yang-Mills fields.

The Skyrmion, however, is unstable to core collapse,
even in the gauged case.
It's core is stabilized in the usual ungauged case
by adding the ``Skyrme term'' which must
be viewed as short-distance
correction to the action:
\be
S_0 = \frac{1}{32}\Tr{([U^\dagger \partial_\mu U, U^\dagger \partial_\nu U])^2}
\ee
In the present case there are indeed  
gauge invariant generalizations of the Skyrme-term,
\be
S_1 = \frac{1}{32}\Tr{([U^\dagger [D_\mu, U], U^\dagger [D_\nu, U]])^2}
= \frac{1}{32}\Tr{([[D_\mu, U^\dagger],[D_\nu, U]])^2}
\ee 
The
gauge invariant Skyrme term $S_1$, with
positive coefficient in the energy, stabilizes the solution
on distance scales $\delta \sim 1/v$. 

The stable Skyrmion solution necessarily involves
the nontrivial near-zone gauge
field configuration in the core. These can be seen
to be identical to the short-distance core of a BPS monopole
\cite{zachos}. 
The chiral field $\phi/v$ is identified with the Wilson
line $\int dx^4 A_4$ and we thus choose the anzatz:
\be
\int dx^4 A_4 \sim \phi/v = f(r)\hat{x}\cdot\vec{\tau}
\ee
For the vector potential we choose:
\be
A^a_i \frac{\tau^a}{2} = 
\frac{h(r)}{g}\vec{x}_j\epsilon^{ijk}\tau_k  
\ee
The energy ($(-1)\times$action for static configurations) then takes the form:
\bea
\label{energy}
E & = & \frac{4\pi}{g^2}\int_0^\infty dr \left[ r^2 \left(h'(r)+\frac{h}{r}\right)^2 
+ \half r^2\left( h^2(r)-\frac{2h}{r} \right) \right]
\nonumber \\
& & + \half v^2 \int_0^\infty dr \left[ r^2 (f'(r))^2 + 2(H(r))^2\sin^2(f(r))\right]
\eea
where it is convenient to introduce the
combination:
\be
\label{Hofr}
H(r) = 1-rh(r)
\ee
Note that $h(r)=2/r$ is a pure gauge configuration.
If we substitute any particular anzatz into
eq.(\ref{energy}) we obtain:
\be
\label{energy2}
E =  \frac{4\pi}{\lambda g^2} c_0 +  \half v^2 \lambda c_1 
\ee
where $c_0$ and $c_1$ are determined from the Yang-Mills
and Skyrmionic  energies respectively.
The energy of the particular anzatz is then relaxed to the
minmum of eq.(\ref{energy2}) with the choice:
\be
\label{energy3}
\lambda^2 = \frac{8\pi c_0}{g^2 v^2 c_1}\;\;\;(a);
\qquad
E = \frac{2v}{g}\sqrt{2\pi c_0c_1}\;\;\;(b)
\ee 
The energy is 
equipartitioned between the two terms of eq.(\ref{energy2}),
which accounts for the factor of $2$ in eq.(\ref{energy3}.b).

To verify that a nontrivial
Yang-Mills field is part of
a stable solution we check that it is
required for binding. We can compare to the energy of the same
Skyrme profile in the case that the Yang-Mills field is switched off:
\bea
\label{energy4}
E_{off} & = & \half v^2 \int_0^\infty dr \left[ r^2 (f'(r))^2 
+ 2\sin^2(f(r))\right]  \equiv \half v^2 c_2
\eea
Thus, we must have:
\be
\frac{E}{E_{off}} < 1, \qquad \makebox{or,}
\qquad \frac{c_1}{c_2} < \half.
\ee
Various choices of  anzatze for the GBS have
been explored numerically.
One is inspired from the
instantonic monopole.  Matching $f(r)$ to $\int dx^4 A_4$ and $h(r)$
to the $x^4=0$ behavior of $A_i$ we obtain:
\be
\label{guess1}
f(r) = \frac{\pi r}{\lambda^2 + r^2},
\qquad h(r) = \frac{2 r}{\sqrt{\lambda^2 + r^2}},
\ee
In fact, we find that this anzatz 
{\em is not bound}, and numerically $E/E_{off} = 1.4$,
not close to a binding a solution. 
The reason is as follows;
we can easily see that for small $r$, and $f(r)$,
our action is equivalent to that of a BPS monopole
(e.g., see the analysis of \cite{zachos}).
The core structure of 
the previous anzatz is far from that of a
BPS monopole.  After some numerical experimentation we
are led to the following:
\be
\label{guess1}
\tilde{f}(r) = \frac{\pi \sqrt{r}}{\sqrt{\epsilon \lambda + r}},
\qquad \tilde{h}(r) = \frac{2 }{\lambda + r},
\ee
Let us initially choose $\epsilon = 1$.
Then we find:
\bea
\tilde{c}_1 & = & \int_0^\infty dx \left[ \frac{\pi^2 x}{4(1+x)^3} 
+ 2\sin^2\left(\frac{\pi \sqrt{x}}{\sqrt{1 + x }}\right)
\left(\frac{1- x}{1 +
x } \right)^2 \right]
\nonumber \\
\tilde{c}_2 & = & \int_0^\infty dx \left[ \frac{\pi^2 x}{4(1+x)^3} 
+ 2\sin^2\left(\frac{\pi \sqrt{x}}{\sqrt{1 + x }}\right) \right]
\eea
and this leads to a net binding:
\be
\frac{E}{E_{off}} = \frac{2\tilde{c}_1}{\tilde{c}_2} = 0.883 < 1
\ee
While the form of eq.(\ref{energy2}) suggests stability of the core
of a solution supported by the Yang-Mills field, generally
we find that the Skyrme profile can be
deformed to collapse and reduce the energy in
the absence of the Skyrme term $S_1$. 
We can, for example, deform the above solution
by choosing $\epsilon \neq 1 $.
We find that the energy is reduced, and the Skyrme
core is unstable. 

The Skyrme term can be added and takes the form
in an anzatz:
\bea
\label{skyrme1}
S_1 & = & {2\pi}\int_0^\infty dr \left[ \sin^2(f)H^2(r)
\left( 2(f')^2 + \frac{1}{r^2} \sin^2(f)H^2(r)  \right) \right]
\eea
$S_1$ must enter the energy with
a positive coefficient and
always dominates at extreme core collapse. The result is
a stable object with a mass given by eq.(\ref{energy3}) of
order $v/g \sim M_{KK}/\alpha$, as in the case of the IM. 

The original conformal invariance of the IM is lost
with the GBS. 
One can also examine
duality and see that it, too, has become only approximate. 
These are no doubt a consequence of compactification \cite{yaffe}
and the truncation on the extreme low energy physics.
The Skyrme
term is a measure of the truncation of theory; the effective deconstructed
theory is not expected to match the short-distance physics
but to capture only the large distance aspects. It would
be instructive to evaluate the Skyrme term
coefficient, as well as the effects
of other operators, such as:
\be
{\cal{O}}^\pm = ig\Tr{F_{\mu\nu}((D^\mu U^\dagger)(D^\nu U)
\pm (D^\mu U)(D^\nu U^\dagger)) }
\ee
which are
analogues of the new currents
of eq.(\ref{cur2}) and eq.(\ref{cur3}).

\newpage
\section{Spontaneously Broken $SU(2)$}

We can add terms to the Lagrangian that are consistent
with the $SU(2) $ symmetry of the form $\sum_p c_p(\Tr(U))^p + h.c.$.
Indeed, such terms must arise at the quantum level, as
in a computation of the 
Coleman-Weinberg potential (see ref.\cite{georgi}). We 
presently add them by hand.
With such terms  we can then destabilize the vacuum;
$\phi$ becomes a Higgs-field which breaks $SU(2)\rightarrow U(1)$,
(as in the recent model of ref.\cite{georgi}, though we do not
presently want an $I=\half$ Higgs).
This means that an arbitrary
VEV of $\phi$ can be engineered,
$\VEV{\phi}=(1-\epsilon)v$ 
where $\epsilon\neq 0$, and is not gauge equivalent
to the unbroken vacuum. 

Since  $\phi$ is an isovector field we have all
of the conditions required for a nontrivial $\Pi_2(SU(2)/U(1))$.
In this case the 
the Gauged-Bosonic Skyrmion grows into a 't Hooft-Polyakov
monopole. The monopole charge is measured by a Chern-Simons charge
in one less dimension, integrated over
the surface at infinity. This contains the dual of $F_{ij}$,
e.g., $\half\Tr\tau_k \cdot \epsilon^{ijk}F_{ij}$ which is integrated over
the surface $d^2\Sigma$
at infinity. We have:
\be
\left. \int_0^\infty r^2\sin\theta d\theta d\phi \;\tilde{F}_r \right|_{r^2\rightarrow
\infty} = 4\pi r^2\frac{H^2-1}{2gr^2} = -\frac{2\pi}{g}
\ee
where $H(r)$ is defined in eq.(\ref{Hofr}), and we see
that asymptotically $H(0)=0$ and $H(\infty)=1$ \cite{zachos}.  
The Skyrme terms now play no
significant role in the core stability since the nontrivial
potential is determining the field value at infinity
and it costs energy to shrink the core. 

Remarkably, however, we see that our monopole is
nontrivially charged under the original $3+1$ Chern-Simons
charge
$\widetilde{Q}$ as well. Including the
gauge degrees of freedom in the Chern-Simons current
in $3+1$ we find that the Chern-Simons charge density
is an exact diffential (see ref.\cite{jackiw}) and the result:
\be
\label{eqxx}
\int d^3x \frac{1}{24\pi^2}\tilde{Q}_0
= \frac{1}{2\pi}[2(f(\infty) -f(0)) - \sin( 2f(\infty))H(\infty)^2
+\sin( 2f(0))H(0)^2]
\ee
Note that no manipulations involving the Chern-Simons
current rely upon the use of equations of motion. 
The monopole anzatz for $f(r)$ is similar
to the Gauge Bosonic Skyrmion, with $f(0) = 0 $ but
now the asymptotic value $f(\infty ) = (1-\epsilon)\pi
\equiv \theta \neq
\pi$. The Chern-Simons charge is now an arbitrary fractional quantity,
$ (\theta - \sin(\theta)\cos(\theta))/\pi$, a result
obtained previously for fermion fractionalization by
Goldstone and Wilczek [13].  The reason is that by forcing $f(\infty )$
to a value less than $\pi$ we  
only partially, map $SU(2)$ into the 3-volume.
Essentially, some fraction
of the Skymion's charge has flowed out to infinity
as the field relaxes into it's nontrivial VEV.

\section{Conclusions} 

We have explored the dimensional descent of a pure Yang-Mills
gauge theory in $4+1$ dimensions, via deconstruction, into
a $3+1$ effective low energy description. We have seen
that topologically nontrivial objects, Instantonic Monopoles,
exist in $4+1$ with nontrivial conserved charges. Under deconstruction
these objects morph into Gauged Bosonic Skyrmions, carrying
a conserved gauge Chern-Simons charge. The scale
size and precise masses are determined by the compactification
scale. The masses of these
objects are $\sim M_{KK}/\alpha$
with core scale sizes $\sim 1/M_{KK}$.  
With spontaneous symmetry breaking, the GBS's further morph into
`t Hooft--Polyakov monopoles. The latter objects carry the
usual magnetic charge, i.e., the magnetic
flux crossing the surface at infinity, 
as well as a fractional Chern-Simons
charge in $3+1$, measuring the partial mapping
of $SU(2)$ into the 3-volume.

All of this 
occurs with pure $SU(2)$ Yang-Mills theory
(it is imbeddible into $SU(N)$) with no explicit Higgs fields,
or explicit chiral fields!  It is a consequence of 
dimensional  compactification, and deconstruction,
which requires the latticization 
of the extra dimensions to maintain the explicit 
manifest gauge invariance.  It is an explicit demonstration
of the descent cohomology of the classical topological solutions 
themselves. Moreover, such objects appear to be
a necessary consequence of Yang-Mills gauge theories propagating
in the bulk with periodic boundary conditions. 

\section*{Acknowledgements}   
   
I wish to thank W. Bardeen and C. Zachos for essential discussions and
for explicitly checking a number of these results. We also thank
J. Goldstone and R. Jackiw for useful comments.   
Research was supported by the U.S.~Department of Energy   
Grant DE-AC02-76CHO3000.

\vspace*{1.0cm}   
\newpage   
 
\frenchspacing   
   
\end{document}